\documentclass[a4paper,11pt]{article}
\usepackage{pos}
\usepackage{amsthm,physics,subcaption}

\title{The Deep Learning Cosmic Ray Energy Reconstruction Pipeline for the GRAPES-3 Experiment}
\ShortTitle{DL CR Energy Reconstruction with GRAPES-3 Experiment}

\author*[a]{Sambit Sarkar}
\author[a]{Mansi Talwar}
\author[a]{Pravata K. Mohanty}
\onbehalf{on behalf of the GRAPES-3 Collaboration}

\affiliation[a]{Tata Institute of Fundamental Research,\\
  Homi Bhabha Road, Mumbai 400005, India}

\emailAdd{sambit.sarkar@tifr.res.in}
\emailAdd{pkm@tifr.res.in}

\abstract{The mass independent energy reconstruction of cosmic rays is crucial for understanding their origin, acceleration, and propagation. Precise measurement of the primary energy can also lead to better mass classification and could enable energy dependent anisotropy maps for individual elements. The GRAPES-3 experiment located in Ooty consisting of 400 scintillator detector array placed 8 m apart covering an area of 25000 m$^2$ with a dedicated muon detector made of 3712 proportional counters, is designed to do these kinds of measurements. Previously electron size calibration curves have been used to find primary energy in the GRAPES-3 data analysis framework however significantly better precision can be established using graph neural network. Thus, in this work we have implemented a modular and dynamic GNN based reconstruction algorithm that automates feature mapping. We demonstrate how the model is learning by studying its latent space and show that scaling the metric in the latent space can lead to further improvements in response resolution. Fine-tuned strategies are presented and a thorough comparison of the reconstructed energy and bias is done for different fine-tuned models along with studying the resolution variation for different mass groups and shower age.}

\ConferenceLogo{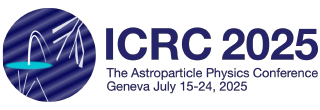}

\FullConference{39th International Cosmic Ray Conference (ICRC2025)\\
 15–24 July 2025\\
Geneva, Switzerland\\}
\begin{document}
\maketitle
\section{Introduction}
\begin{figure}[t]
    \centering
    \begin{subfigure}{0.45\linewidth}
    \centering
        \includegraphics[width=1\linewidth]{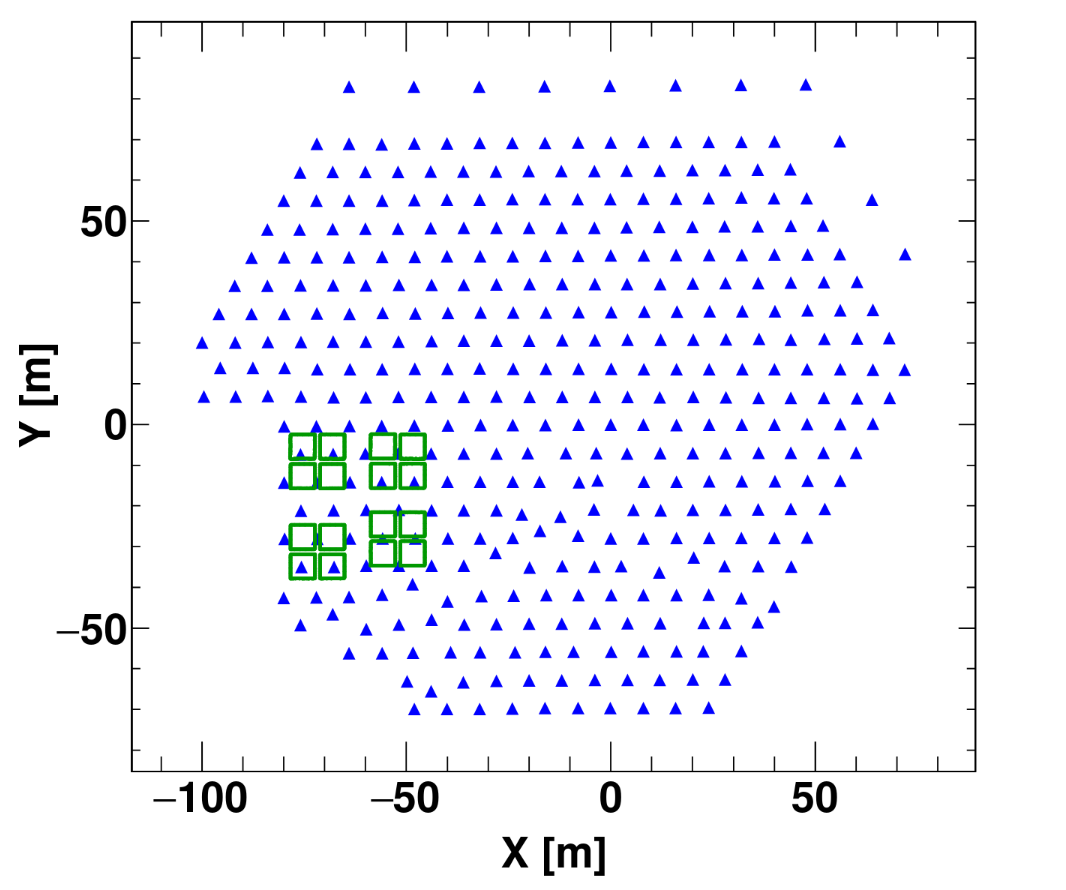}
        \caption{}
        \label{fig:arrayimg}
    \end{subfigure}
    \quad
    \begin{subfigure}{0.45\linewidth}
    \centering
        \includegraphics[width=1\linewidth]{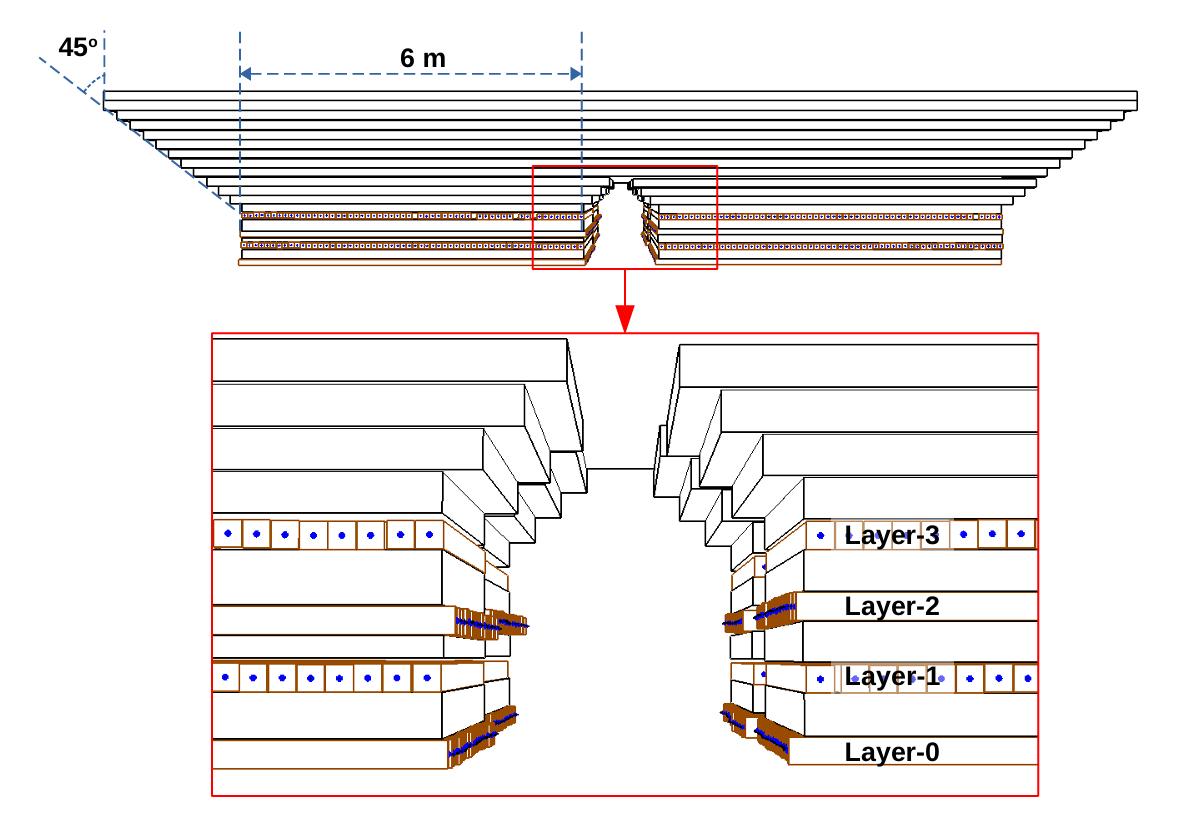}
        \caption{}
        \label{fig:gridsim}
    \end{subfigure}
    \caption{(a) The GRAPES-3 detector array where blue and green rectangles give the position of scintillator detectors and the G3MT modules respectively (b) Schematic of a G3MT module: top shows its cross-section; bottom illustrates the four PRC layers with alternating alignment \cite{Varsi2023}.}
    \label{fig:array}
\end{figure}

The GRAPES-3 (Gamma Ray Astronomy at PeV EnergieS - phase 3) is a ground-based extensive air shower (EAS) experiment located at Ooty in India (11.4$^\circ$N, $76.7^\circ$E, 2200 m a.s.l). It consists of a dense array of nearly 400 plastic scintillator detectors (1 m$^2$ each), spaced 8 m apart, to record the density and arrival time of EAS secondary particles. With an effective area of 25,000 m$^2$, GRAPES-3 achieves a low energy threshold of 100 TeV, overlapping with direct measurements. The muon content is measured by the 560 m$^2$ muon telescope (G3MT), comprising of 3712 proportional counters arranged in 16 modules, each with 35 m$^2$ area having four layers, with 58 PRCs per layer. A single PRC is 600 cm long, consists of a galvanized steel tube with 100 cm$^2$ square cross-section and 0.23 cm wall thickness, sealed at both ends with 0.6 cm thick steel plates. It contains P-10 gas and a 100 $\mu$m tungsten anode wire along its axis. The layers are alternately aligned east-west and north-south to optimize muon track reconstruction in the Y-Z and X-Z planes. Concrete blocks of dimension 60 $\times$ 60 $\times$ 15 cm$^3$ are used between adjacent layers to only allow muons and hadronic tracks \cite{Varsi2023}. The detector setup is schematically shown in \autoref{fig:array}. Among its achievements, GRAPES-3 has observed a spectral hardening near 166 TeV for protons \cite{Varsi2024}, confirmed small-scale cosmic ray anisotropy at TeV energies \cite{Chakraborty2024}, and made strides in gamma-ray classification and astronomy \cite{Dipti}.

So far however, a dedicated algorithm for energy reconstruction of cosmic rays is absent from the GRAPES-3 pipeline. Precise energy reconstruction is important as it enables finer resolution of the cosmic ray spectrum, improves mass reconstruction, and supports energy-dependent anisotropy studies. Graph neural networks (GNNs) have recently shown success in reconstructing the primary energy of charged pion showers in the CMS high-granularity calorimeter
\cite{Aamir2024}. Motivated by this, we developed a GNN-based model to reconstruct the primary energy of cosmic rays using both detector and high-level features from GRAPES-3 EAS simulations. In this work, we present our model discussing how it works and the resolution and bias it can achieve with a fine-tuning strategy.

\section{Scaling Dynamic Reduction Network}\label{GNN}
\begin{figure}[t]
    \centering
    \includegraphics[angle=90,width=0.7\linewidth]{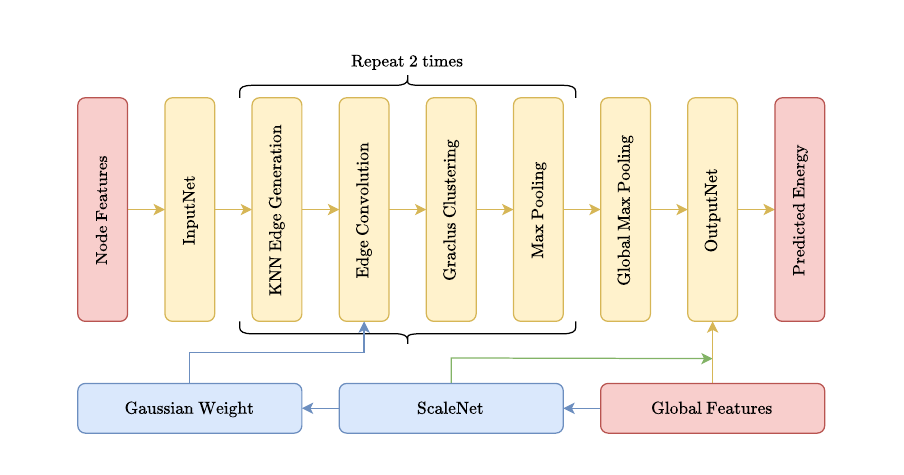}
    \caption{Schematic of DRN and its modification, where the red blocks denote the input and output variables, yellow blocks denote the core DRN networks and processes, blue blocks denote SDRN modifications and green line denotes fine-tuning modification.}
    \label{fig:DRNmodel}
\end{figure}

Typical GNN node feature mappings significantly impact target precision and labeled cluster separation. Let $\mathcal{P}$ be the input manifold and $\mathcal{L}$ the latent space where node distances are maximally sensitive to primary energy. We define the MLP-based transformation $I:\mathcal{P} \to \mathcal{L}$ as \verb|InputNet|, with $\dim(\mathcal{L}) > \dim(\mathcal{P})$ to enhance nearest-neighbor and class separability. To increase node sensitivity within local clusters (identified by Graclus), we apply a permutation-invariant message-passing MLP $E:\mathcal{L} \to \mathcal{L}'$ (e.g., edge convolution), with $\dim(\mathcal{L}) = \dim(\mathcal{L}')$. Max pooling selects the most sensitive node per cluster, reducing node count. Iterating clustering, convolution, and pooling enables hierarchical feature learning, typically requiring two steps—three often degrades energy resolution. The final representation is mapped to the energy manifold $\mathcal{E}$ via \verb|OutputNet| $O:\mathcal{L}' \to \mathcal{E}$, optionally incorporating global features. This architecture, known as the Dynamic Reduction Network (DRN), has proven effective for point-cloud-based learning in extensive air showers \cite{gray2020dynamicreductionnetworkpoint}. The complete schematic of DRN and its variants is shown in \autoref{fig:DRNmodel}.

\begin{figure}[!t]
\centering
\begin{subfigure}{0.45\textwidth}
    \centering
    \includegraphics[width=1\linewidth]{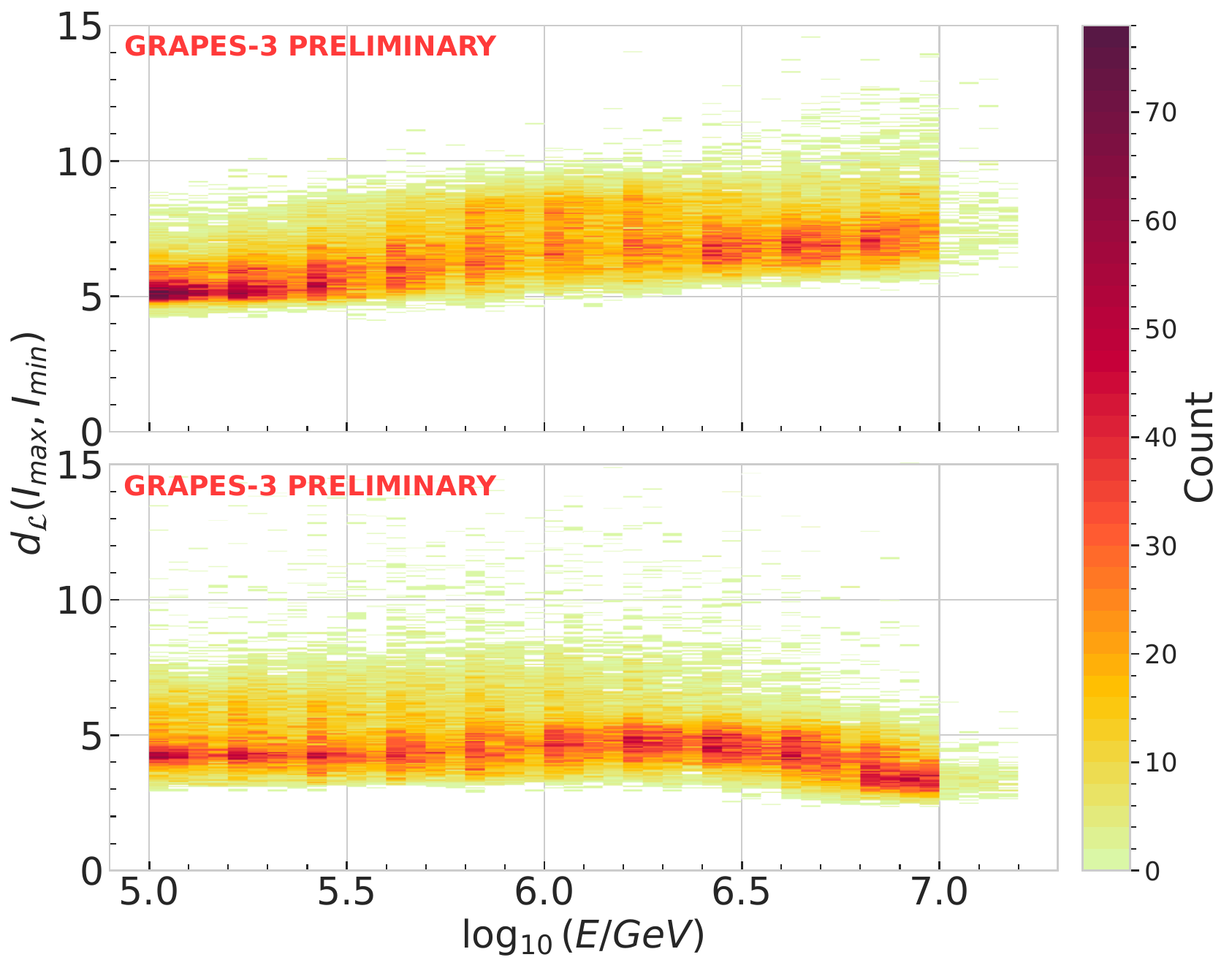}
    \caption{}
    \label{fig:Ereldist}
\end{subfigure}
\quad
\begin{subfigure}{0.451\textwidth}
    \centering
    \includegraphics[width=1\linewidth]{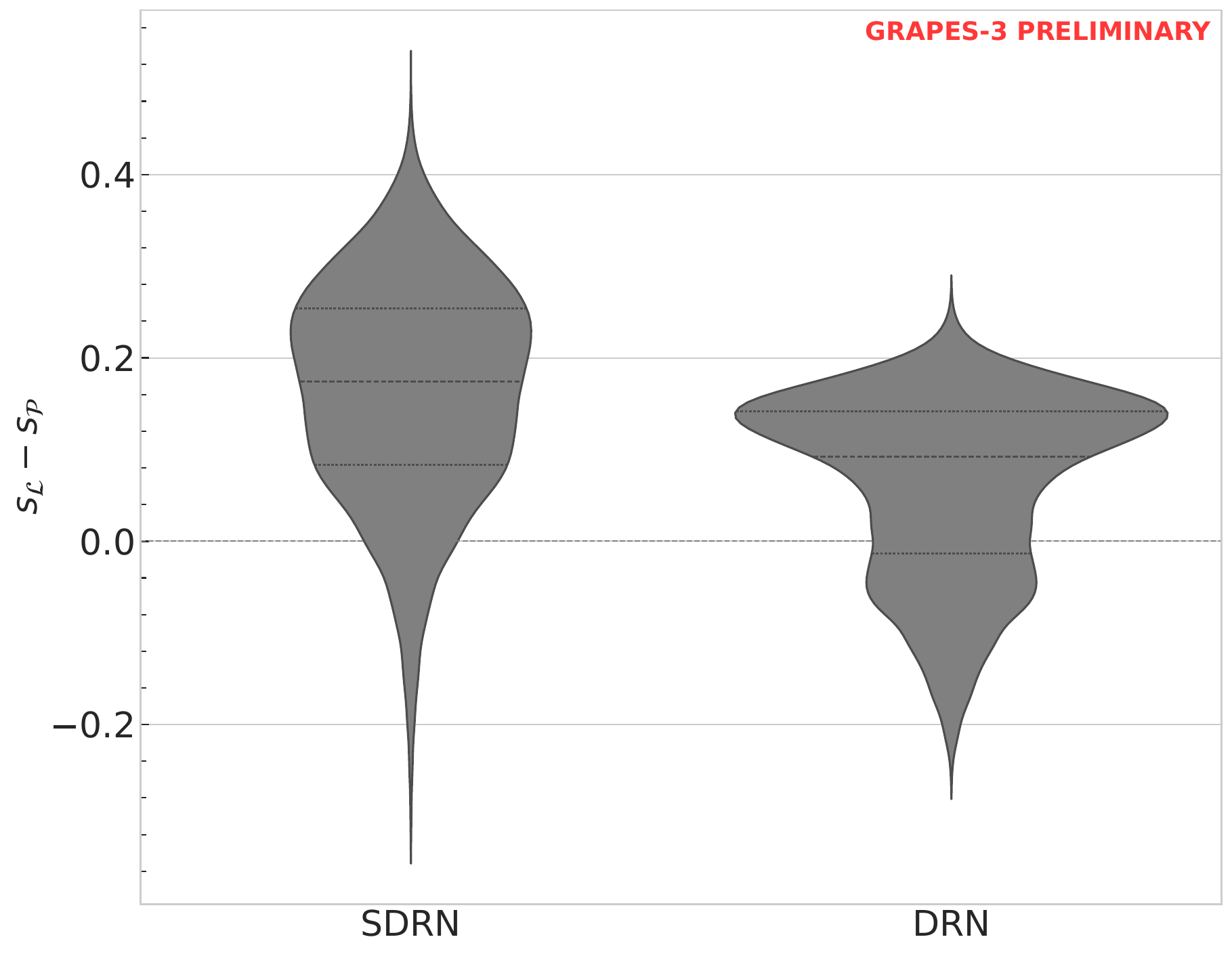}
    \caption{}
    \label{fig:sviolin}
\end{subfigure}
\caption{(a) The latent length variation with log energy where (top) plot denotes DRN and (bottom) denotes SDRN. (b) The violin plot showing improvement in G3MT module node separation from SD nodes for the SDRN.}
\end{figure}

The transformation $I$ is not distance-preserving in DRN, leading to extrusion at higher energies and introducing significant bias in energy representation. To address this, we manipulate the metric so that embeddings in the latent space $\mathcal{L}$ becomes wrapped ensuring node distances do not scale with energy. This is achieved during edge convolution by applying a distance-based weight to the message-passing MLP. While such weights can be empirically estimated \cite{Koundal:2023F}, we introduce an MLP called \verb|ScaleNet| $(S)$,which takes global features as input and predicts a scaling parameter $\kappa$, from which a Gaussian weight $w_{ij}$ is computed as,
\begin{equation}
    w_{ij} = \exp[-\frac{d_{\mathcal{L}}(I(x_i),I(x_j))^2}{\kappa^2}]
\end{equation}
The scaled weight is applied to each edge message, reflecting its importance. This model is referred to as the scaling dynamic reduction network (SDRN). To evaluate its impact, we track how the distance between the closest and farthest nodes in $\mathcal{L}$—denoted $I_{{min}}$ and $I_{{max}}$—varies with energy. Their separation, $d_{\mathcal{L}}(I_{\text{max}}, I_{\text{min}})$, defines the latent length (LL) of a shower. As energy increases, more SDs are hit, increasing the node count. If LL remains fixed, this implies shorter effective nearest-neighbor distances. This is essential for edge convolution, as it prevents low-energy events with sparse neighbors from resembling high-energy events where LL has scaled up. In \autoref{fig:Ereldist}, the DRN (top) shows increasing LL with energy, while SDRN (bottom) maintains a more stable LL via metric scaling. The slight drop in LL near 10 PeV is discussed later. Notably, SDRN shows a marked improvement in LL resolution around 1 PeV over DRN, which also translates to better energy resolution, as seen in \autoref{fullmodeltrain}. Another key test for validating energy representations in $\mathcal{L}$ is the separation between detector types. SDs capture the overall shower profile, while G3MT is sensitive to both energy and mass composition. We quantify this separation using the silhouette score, defined as,
\begin{equation}
    s_{\mathcal{M}} = \expval{\frac{b_{\mathcal{M}}(i)-a_{\mathcal{M}}(i)}{\max\qty[a_{\mathcal{M}}(i),b_{\mathcal{M}}(i)]}}
\end{equation}
where $a_{\mathcal{M}}(i)$ is the average distance between node $i$ and other nodes in the same cluster in the manifold $\mathcal{M}$ while $b_{\mathcal{M}}$ is the lowest average distance to nodes in other clusters. The average is taken over all nodes for a given shower event. A silhouette value of 1 indicates perfect separation, 0 implies overlapping clusters, and negative values suggest misclustering.  \autoref{fig:sviolin} shows violin plots of $s_{\mathcal{L}} - s_{\mathcal{P}}$ for DRN and SDRN, with dotted lines marking the median and quartiles. DRN shows misclustering in the first quartile, while SDRN improves intra-cluster consistency with first quartile > 0, and class separation with third quartile > 0.2.

\begin{figure}[!t]
\centering
\begin{subfigure}{0.425\textwidth}
    \centering
    \includegraphics[width=1\linewidth]{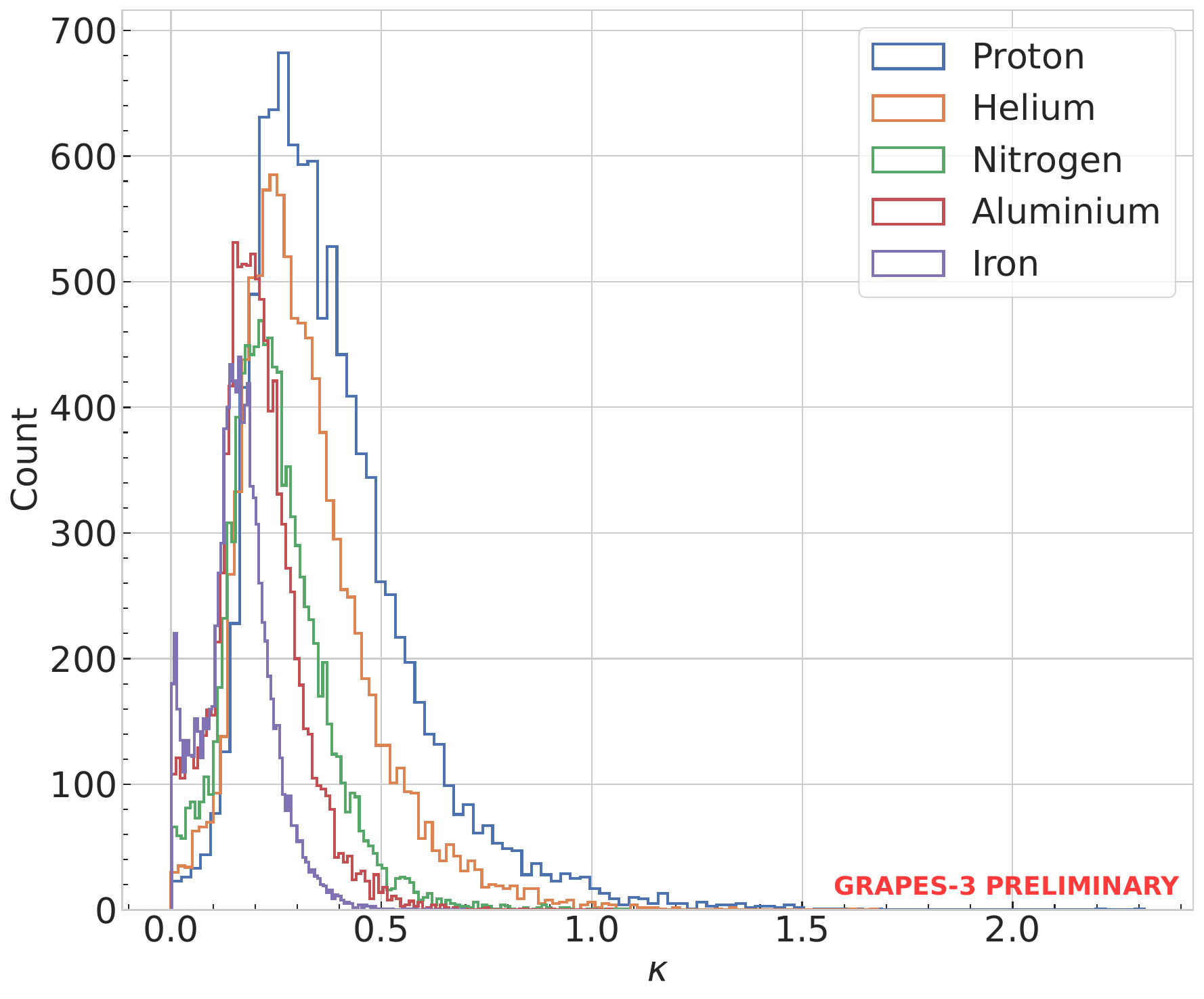}
    \caption{}
    \label{fig:kmasshist}
\end{subfigure}
\quad
\begin{subfigure}{0.418\textwidth}
    \centering
    \includegraphics[width=1\linewidth]{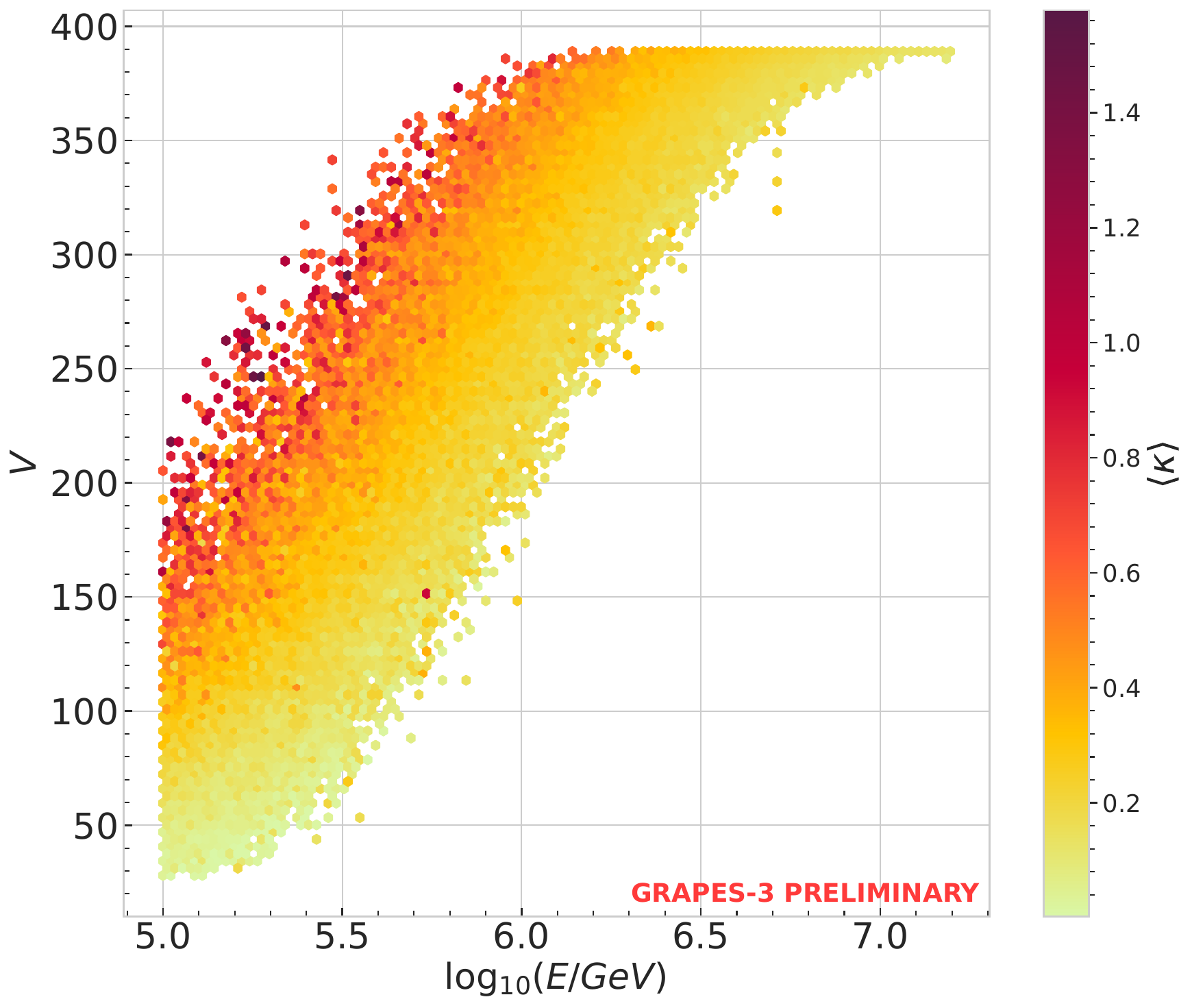}
    \caption{}
    \label{fig:VEnergyScale}
\end{subfigure}
\caption{(a) Mass dependent histograms of $\kappa$ where we can see separate distributions for the five mass groups with iron (proton) having the smallest (largest) variance (b) Variation of $\expval{\kappa}$ with the total detector hits show that for any given energy a higher detector hit would correspond to a larger average scale.}
\end{figure}

An interesting outcome of LL-based scaling is that the scaling parameter $\kappa$ becomes mass-sensitive, as shown in \autoref{fig:kmasshist}, since lighter primaries with larger shower maximum produce showers with more charged particles, increasing LL. \autoref{fig:VEnergyScale} shows that $\expval{\kappa}$ grows with both energy and node count $V$, and for fixed energy, larger $V$ implies higher $\kappa$. The deviation from constant LL beyond $\log_{10}(E/\text{GeV}) = 6.5$ can now be attributed to detector saturation, reinforcing the idea that $\kappa$ controls effective node density. For the models, we used 3-layer \verb|InputNet| and \verb|OutputNet|, 2-layer \verb|EdgeNet| and \verb|ScaleNet|, with $\dim(\mathcal{P})=9$, $\dim(\mathcal{L})=64$, and 16 nearest neighbors for DRN, 32 for SDRN to account for increased metric variation.

\section{Feature Mapping and Validation}\label{fmap}

The node features include detector coordinates (centered on the array), shape of SD and G3MT modules, log-normalized charged particle density (SD) and reconstructed muon and hadron counts (G3MT), log-normalized SD arrival times and average SD arrival time for G3MT due to limited timing resolution, and a detector type label. The global features include Nishimura-Kamata-Greisen (NKG) fitted shower age and size \cite{Varsi2023}, total SDs hit, total reconstructed muons (G3MT),  total charged particle density in SDs atop G3MT (small sample proxy for $n_\mu/n_e$), standard deviation of SD density and timing, and shower core location \cite{Vuta:2023i6}. During fine-tuning $\kappa$ is supplied to the \verb|OutputNet| as well. For this work, we try to predict the energy in PeV scale. To validate the models, we calculate the response as $E_{pred}/E_{true}$, logarithmically binned and fit each distribution with the Cruijff function given as,
\begin{equation}
    f(x|\mu,\sigma_L,\sigma_R,\alpha_L,\alpha_R,N) = N\begin{cases}
        \exp[-\frac{(x-\mu)^2}{2\sigma_L^2+\alpha_L(x-\mu)^2}] & x<\mu\\
        \exp[-\frac{(x-\mu)^2}{2\sigma_R^2+\alpha_R(x-\mu)^2}] & x\geq\mu
    \end{cases}
\end{equation}
Here $\sigma_{L/R}$ capture the left/right standard deviation while $\alpha_{L/R}$ capture left/right tail variations. We use $\mu$ to denote the bias of the response, while $\sigma=(\sigma_L+\sigma_R)/2$ gives the resolution.
\section{Results}\label{fullmodeltrain}
\begin{figure}[!t]
    \centering
    \includegraphics[width=0.5\linewidth]{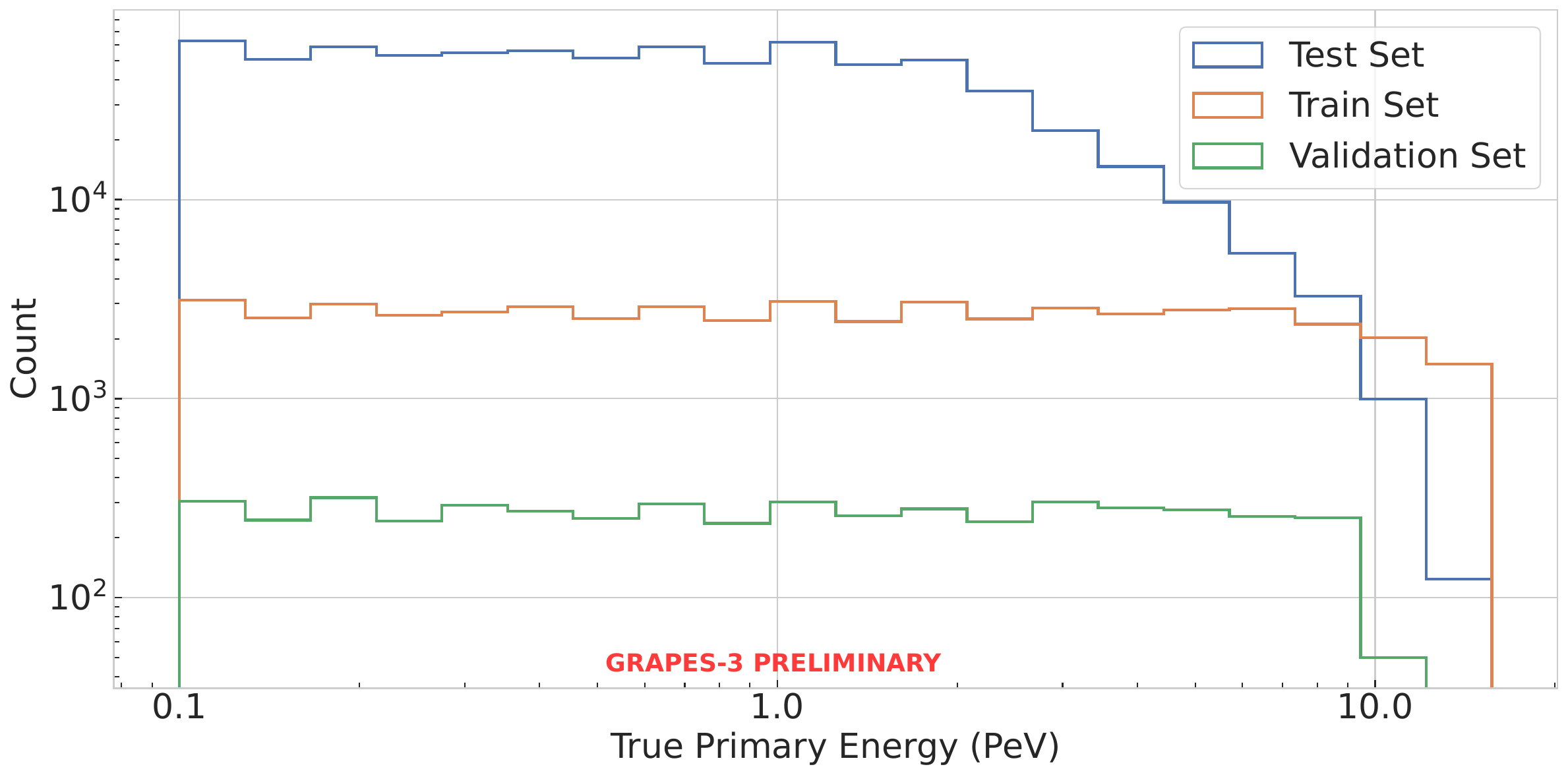}
    \caption{The dataset distribution over the 20 bins considered for model validation.}
    \label{fig:ehist}
\end{figure}
We randomly selected events from the GRAPES-3 dataset with energies following a $-2.5$ spectral index, spanning 100 TeV to 16 PeV in 11 logarithmic bins (width 0.2), generated using CORSIKA 7.69 with QGSJETII-04 and FLUKA for hadronic interactions, and EGS4 for electromagnetic interactions \cite{Varsi2024}. Each shower was reused 10 times with a random core within 60 m of the array center and zenith/azimuth ranges $0^\circ < \theta < 45^\circ$, $0^\circ < \phi < 360^\circ$. Detector responses were simulated in GEANT4, and muon tracks reconstructed for each triggered event. For finer resolution, the energy range was split into 20 logarithmic bins (\autoref{fig:ehist}) with nearly uniform distribution across five mass groups. Proton and helium events beyond 10 PeV were excluded due to detector saturation. Training and testing were performed on a single RTX 4090 GPU.

\begin{figure}[!b]
\centering
\begin{subfigure}{0.45\textwidth}
    \centering
    \includegraphics[width=1\linewidth]{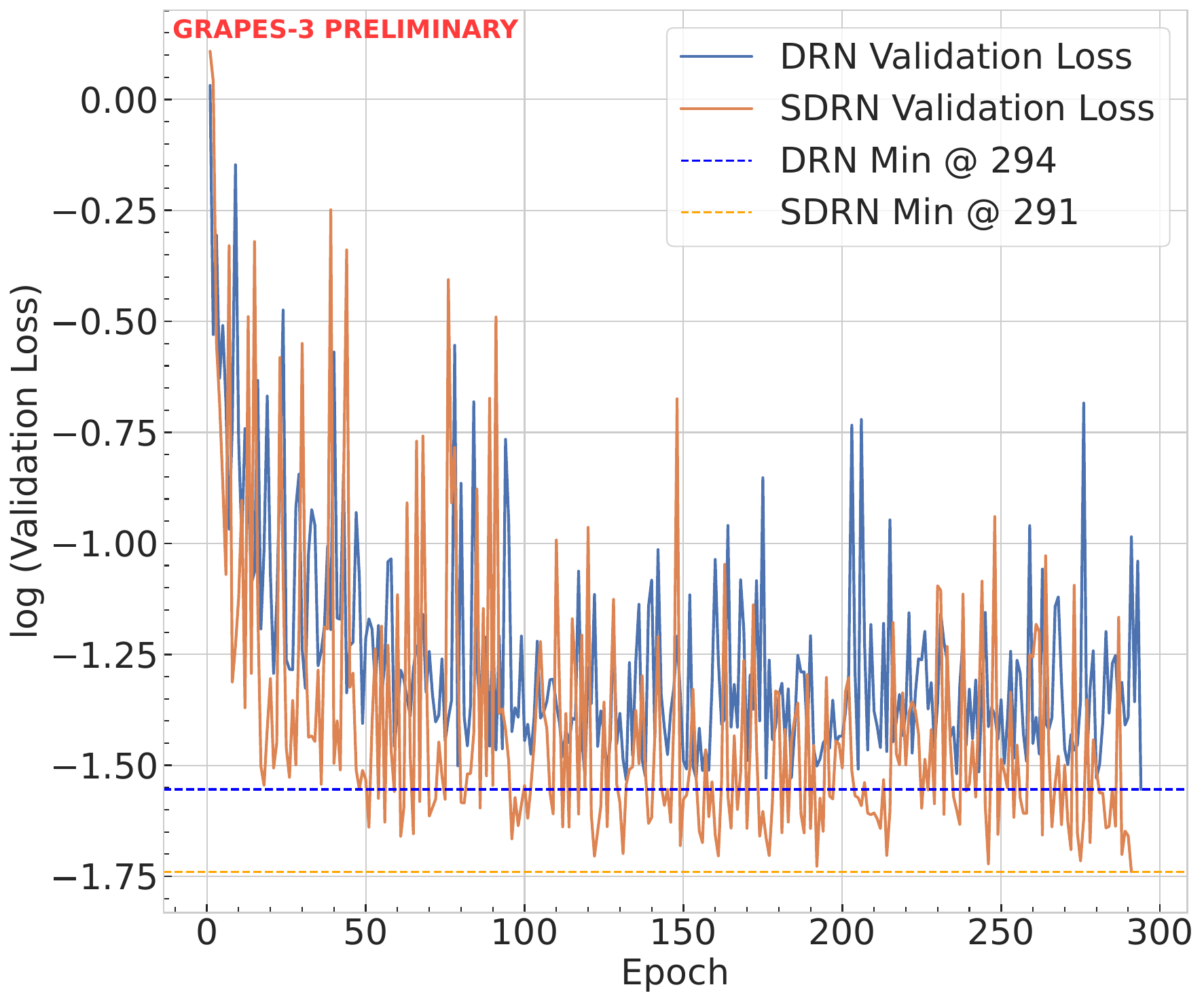}
    \caption{}
    \label{fig:valloss}
\end{subfigure}
\quad
\begin{subfigure}{0.46\textwidth}
    \centering
    \includegraphics[width=1\linewidth]{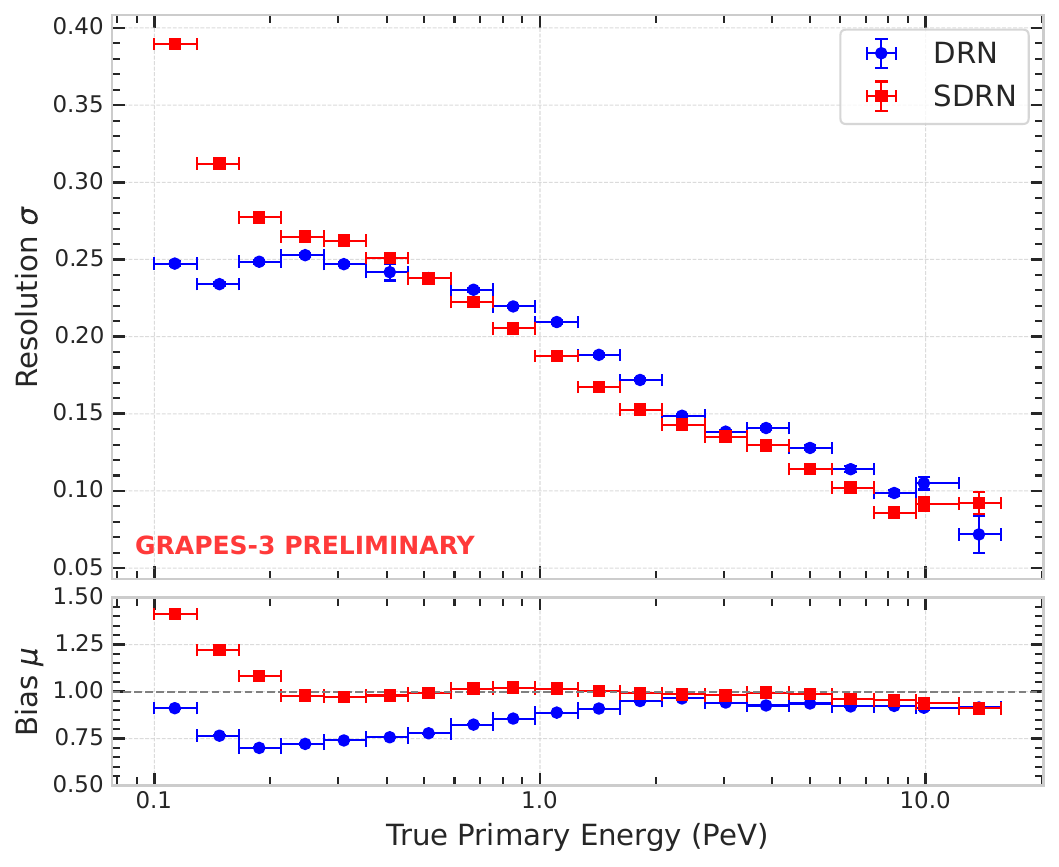}
    \caption{}
    \label{fig:fulltrainres}
\end{subfigure}
\caption{(a) Comparison showing that SDRN is able to achieve a much lower log loss than DRN. (b) The resolution and bias of the energy response for DRN and SDRN.}
\end{figure}

\begin{figure}[!t]
\centering
\begin{subfigure}{0.46\textwidth}
    \centering
    \includegraphics[width=1\linewidth]{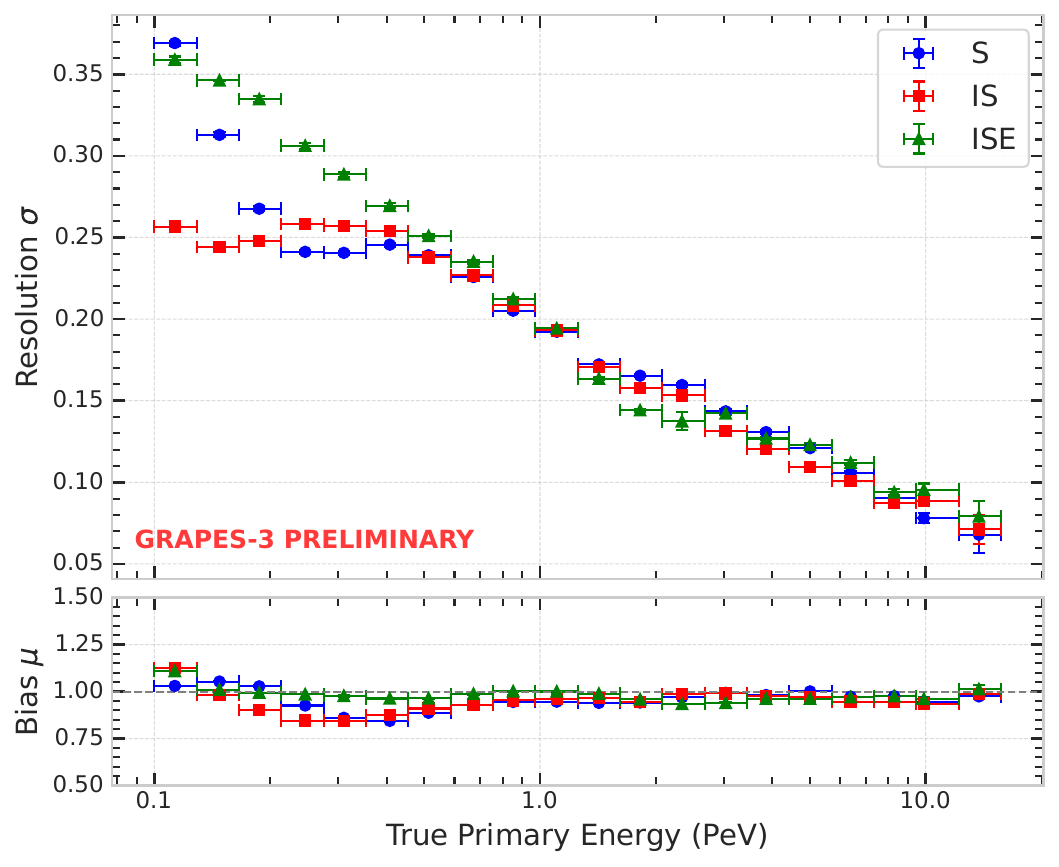}
    \caption{}
    \label{fig:firstFT}
\end{subfigure}
\quad
\begin{subfigure}{0.46\textwidth}
    \centering
    \includegraphics[width=1\linewidth]{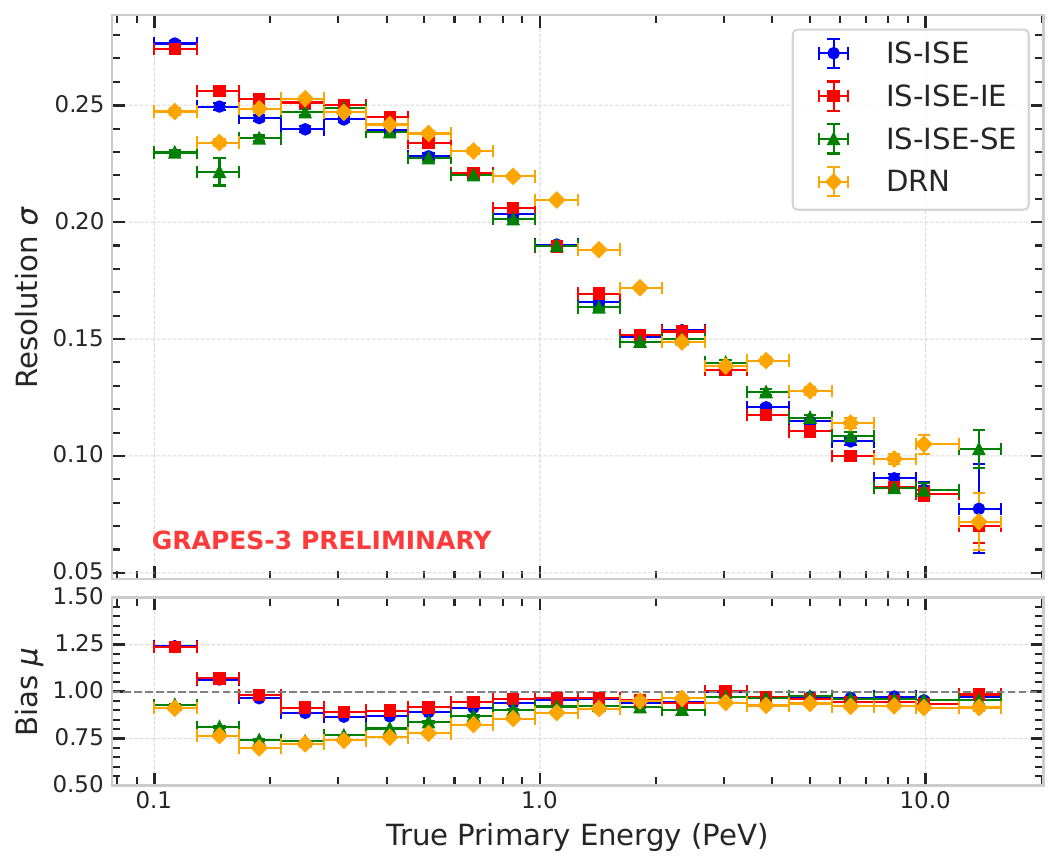}
    \caption{}
    \label{fig:secondFT}
\end{subfigure}
\caption{Comparison of (a) models after first fine-tuning where we see that training only OutputNet removes bias while training the InputNet degrades the resolution (b) models after second fine-tuning with DRN.}
\end{figure}

In figure \autoref{fig:valloss} we can see that SDRN generally has a lower loss however there is significant fluctuation due to initial latent representation introduced by \verb|ScaleNet|. The resolution and bias of the response of the two models is plotted in \autoref{fig:fulltrainres} evidently showing superiority of SDRN at energy higher than 600 TeV upto about 10 PeV, with the highest difference around 1 PeV. The bias on the other hand improves drastically for energy above 200 TeV. However, below this SDRN has poor resolution and bias owing to the large scaling-induced metric variations acting as noise for the \verb|OutputNet|. This can be fixed by attaching $\kappa$ to \verb|OutputNet|. But for the same reason as before, such a model cannot be trained with all weights active. We will therefore fix the pre-trained weights for some of the components and train the rest.

\begin{figure}[!b]
\centering
\begin{subfigure}{0.46\textwidth}
    \centering
    \includegraphics[width=1\linewidth]{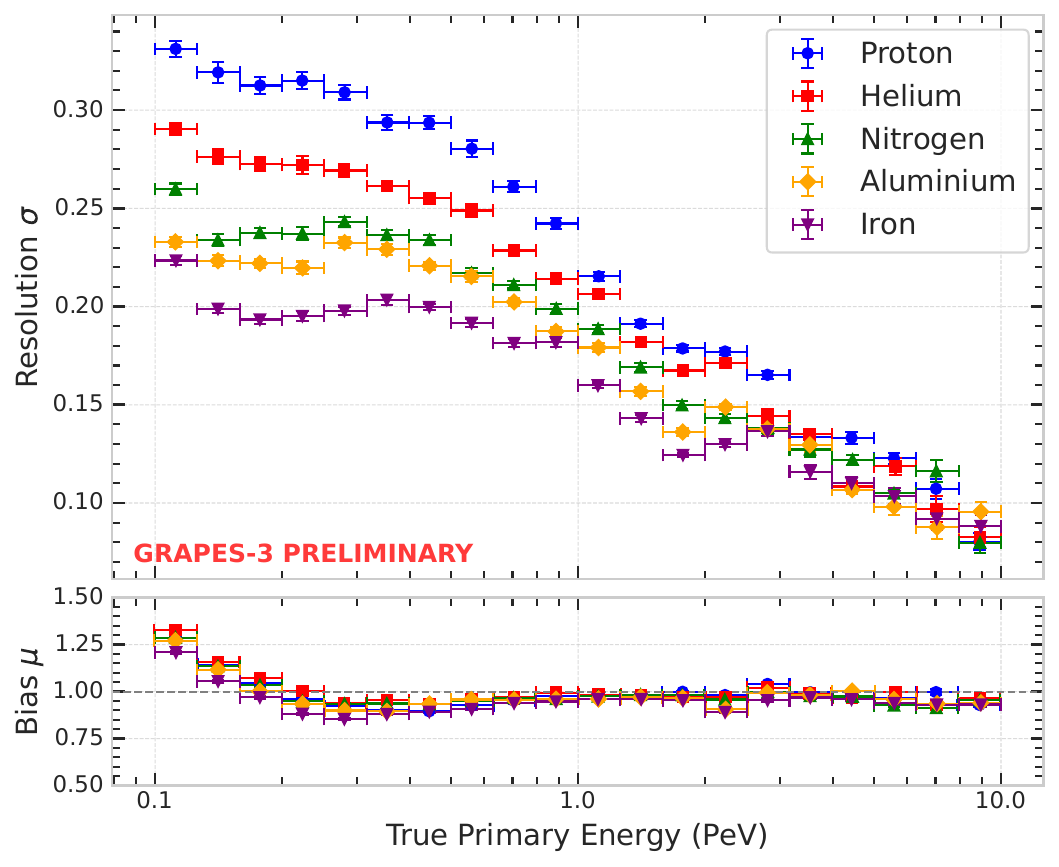}
    \caption{}
    \label{fig:massISISEIE}
\end{subfigure}
\quad
\begin{subfigure}{0.451\textwidth}
    \centering
    \includegraphics[width=1\linewidth]{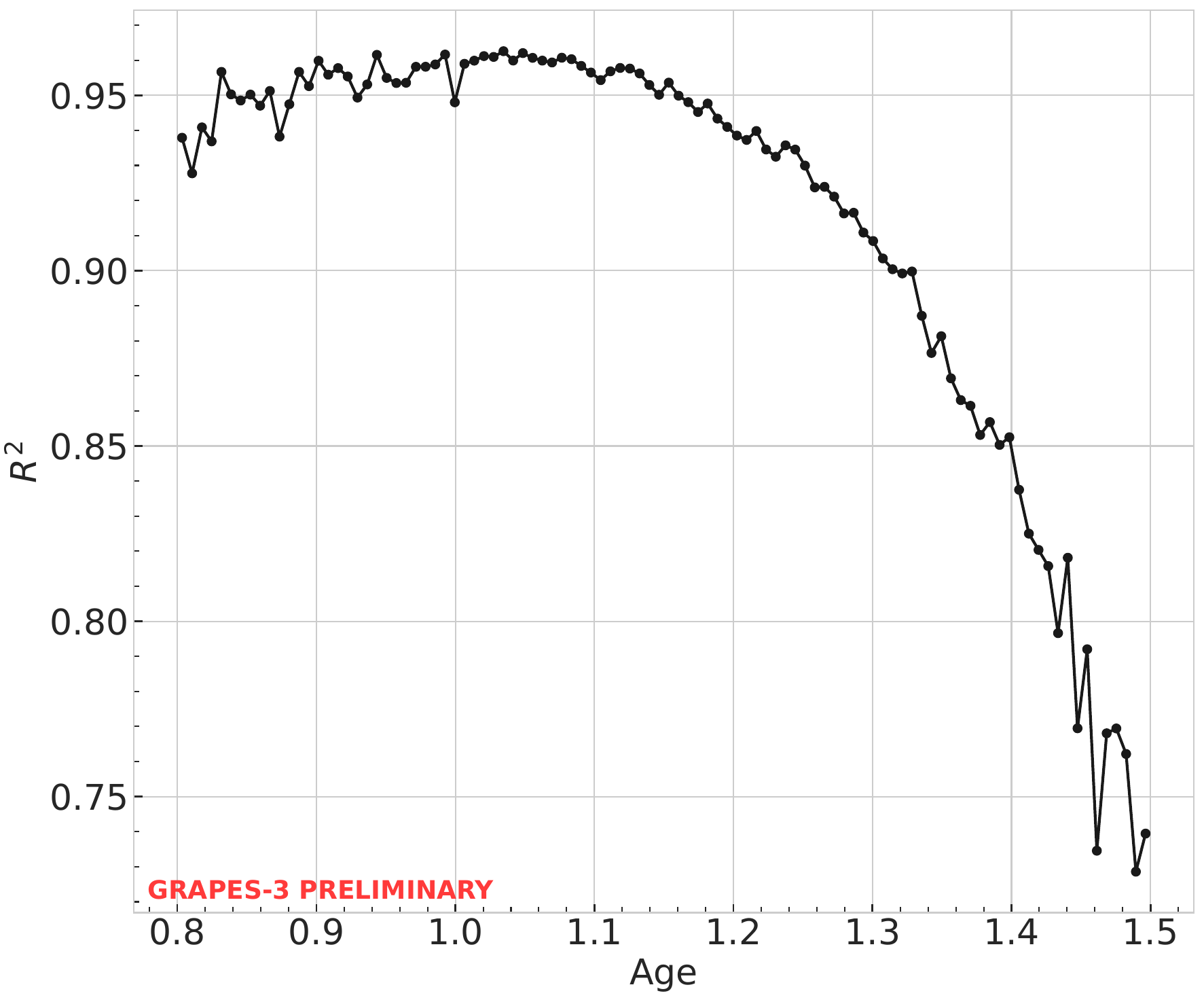}
    \caption{}
    \label{fig:r2age}
\end{subfigure}
\caption{(a) Mass dependent resolution and bias for IS-ISE-IE showing consistency till 2 PeV above which node saturation occurs. (b) Variation of $R^2$ with age is shown for IS-ISE-IE where the best performance is reached around 1.}
\end{figure}

For fine-tuning, we use a naming scheme where fixed components retain their pre-trained SDRN weights; e.g., the ISE model has \verb|InputNet|, \verb|ScaleNet|, and \verb|EdgeNets| fixed. As shown in \autoref{fig:firstFT}, ISE maintains a bias close to 1 across all energies, while IS improves low-energy resolution at the cost of higher bias. This motivates a sequential fine-tuning strategy where \verb|EdgeNet| and \verb|ScaleNet| are tuned along with \verb|OutputNet|. Bias introduced in this process can be mitigated by a final ISE training step. Models following this scheme—like IS-ISE-IE and IS-ISE-SE—are compared with DRN in \autoref{fig:secondFT}. Retraining \verb|InputNet| below 1 PeV leads to instability, often introducing irreducible bias or degraded resolution due to noisy metrics. Hence, we exclude S and IS-ISE-SE models across most energies, except near 100 TeV. Between 200 TeV and 3 PeV, IS-ISE performs best, while above 3 PeV, IS-ISE-IE is preferable. \autoref{fig:massISISEIE} shows the IS-ISE-IE model’s resolution and bias by mass group.  Below 2 PeV, protons have the poorest resolution—60–70\% worse than iron—though biases remain similar. Above 2 PeV, resolution degrades slightly for most groups except proton and nitrogen, likely due to node saturation. At higher energies, latent representations shift, reducing LL’s ability to encode energy and degrading resolution for heavier primaries. As seen in \autoref{fig:r2age}, $R^2$ improves notably near shower maximum, motivating a selection cut of $0.9 < \text{age} < 1.15$ for better reconstruction.

\section{Conclusion and Future Work}\label{conclusion}
We demonstrate that primary cosmic ray energy can be accurately reconstructed using a Dynamic Reduction Network, which maps EAS data to a latent manifold $\mathcal{L}$ where energy is learned via metric representations. Local rescaling of the metric in $\mathcal{L}$ reduces response bias and improves high-energy resolution. We also demonstrated a fine-tuning strategy leading to multiple models that can be combined for a future ensemble learning approach. Further gains can be achieved by incorporating electron and muon counts through a hierarchical model or multi-head \verb|OutputNet|s with GradNorm-weighted loss, constraining the $(E, n_e, n_\mu)$ phase space. Finally, introducing energy bin information during regression may further improve scale invariance. This work establishes the foundation for a deep learning pipeline in GRAPES-3 and enables future enhancements.
%\subsection{Energy Reconstruction Using True Parameters}
%\subsection{Philosophy of Hierarchal Reconstruction}
%\subsection{Constraining the $(E,n_e,n_\mu)$ Phase Space}
%\subsection{Unfolding on the Energy Bins}
%\section{Conclusion}
\acknowledgments
S. Sarkar thanks Shilpi Jain, Rajdeep Chatterjee, Medha Chakraborty, Diptiranjan Pattanaik, Ronald Scaria, Mohan Karthik, Fahim Varsi, and Paras Koundal for valuable comments.

\bibliographystyle{jhep.bst}
\bibliography{bibliography.bib}

%\section{...}

%\begin{thebibliography}{99}
%\bibitem{...}
%....

%\end{thebibliography}

\end{document}